\documentclass[useAMS,usenatbib]{mn2e}
\usepackage{txfonts,graphicx,amssymb,subfigure,color}
\usepackage[normalem]{ulem}
\usepackage{natbib}
\newcommand\HI{H\,{\small I}~}

\usepackage[colorlinks=true, citecolor=blue,linkcolor=black]{hyperref}
\begin{document}

\title[The non-thermal superbubble in IC~10]{The non-thermal superbubble in IC~10: the generation of cosmic ray electrons caught in the act}

\author[V.~Heesen et al.]{Volker Heesen,$^{1}$\thanks{E-mail: v.heesen@soton.ac.uk} Elias
  Brinks,$^{2}$ Martin G.~H.~Krause,$^{3,4}$ Jeremy J.~Harwood,$^{2}$\thanks{Now at
    ASTRON, Postbus~2, 7990 AA Dwingeloo, the Netherlands.} \newauthor Urvashi
  Rau,$^{5}$ Michael P.~Rupen,$^{5}$ Deidre A.~Hunter,$^{6}$
  Krzysztof T.~Chy\.zy$^{7}$ \newauthor and Ged Kitchener$^2$\\
$^{1}$School of Physics and Astronomy, University of Southampton,
Southampton SO17 1BJ, UK\\
$^{2}$Centre for Astrophysics Research, University of
Hertfordshire, Hatfield AL10 9AB, UK\\
$^{3}$Excellence Cluster Universe, Technische Universit\"at M\"unchen, Boltzmannstrasse 2, D-85748 Garching, Germany\\
$^{4}$Max-Planck-Institut f\"ur extraterrestrische Physik, Giessenbachstr. 1, D-85741 Garching, Germany\\
$^{5}$NRAO, P.V.D. Science Operations Center, National Radio Astronomy Observatory, 1003 Lopezville Road, Socorro, NM 87801, USA\\
$^{6}$Lowell Observatory, 1400 West Mars Hill Road, Flagstaff, AZ 86001, USA\\
$^{7}$Obserwatorium Astronomiczne Uniwersytetu Jagiello\'nskiego, ul. Orla
171, 30-244 Krak\'ow, Poland\\}

\date{Accepted 2014 October 15. Received 2014 October 15; in original form
  2014 September 22}

\maketitle

\begin{abstract}
  Superbubbles are crucial for stellar feedback, with supposedly high (of the order of 10 per cent) thermalization
  rates. We combined multiband
  radio continuum observations from the Very Large Array (VLA) with Effelsberg data to study the
  non-thermal superbubble (NSB) in IC~10, a
  starburst dwarf irregular galaxy in the Local Group. Thermal emission was
  subtracted using a combination of Balmer H$\alpha$ and VLA
  32~GHz continuum maps. The bubble's non-thermal spectrum between $1.5$ and
  $8.8$~GHz displays curvature and can be well
 fitted with a standard model of an ageing cosmic ray electron
 population. With a derived equipartition magnetic field strength
  of $44\pm 8~\rm\mu G$, and measuring the radiation energy density from
  \emph{Spitzer} MIPS maps as $5\pm 1
\times 10^{-11}~\rm erg\,cm^{-3}$, we determine, based on the spectral
curvature, a spectral age of the bubble of
  $1.0\pm 0.3$~Myr. Analysis of the LITTLE
 THINGS \HI data cube shows an expanding \HI hole with 100~pc
 diameter and a dynamical age of $3.8\pm 0.3$~Myr, centred to within 16~pc on IC~10 \hbox{X-1}, a
 massive stellar mass black hole ($M>23~\rm M_{\sun}$). 
The results are consistent with the expected evolution for a superbubble
with a few massive stars, where a very energetic event like a Type~Ic supernova/hypernova
has taken place about 1~Myr ago. We discuss alternatives to this interpretation.
\end{abstract}

\begin{keywords}
radiation mechanisms: non-thermal -- cosmic rays -- galaxies: individual:
IC~10 -- galaxies: irregular  -- galaxies: starburst -- radio continuum: galaxies.
\end{keywords}

\section{Introduction}
\label{sec:introduction}
Stellar feedback is a fundamental process that regulates the
formation and evolution of galaxies. Supernovae (SNe) inject energy into the
interstellar medium (ISM), heating the gas to X-ray emitting temperatures and
accelerating cosmic rays via shock waves. Galactic winds, hybridly driven by
the hot gas and cosmic rays, remove mass and angular momentum \citep{everett_08a, strickland_09a, dorfi_12a,
  hanasz_13a, salem_14a}. Cosmological
simulations without stellar feedback not only predict wrong global mass
estimates, but mass concentrations towards the centres of galaxies that are too
high, leading to rotation curves that are steeper than observed \citep{scannapieco_13a}. The most
abundant type of galaxies in the local Universe, dwarf galaxies, are particularly affected by outflows: their weak gravitational potentials
make them susceptible to outflows and winds \citep{tremonti_04a}. In the paradigm of a
$\Lambda$CDM Universe, the removal of baryons in
the least massive dark matter haloes may
resolve the long standing `missing satellites' problem \citep{moore_99a}. The
loss of baryonic matter and associated angular momentum at early stages in
their formation and evolution can affect the distribution of the non-baryonic
matter as well, rendering the inner part of the rotation curves less
steep \citep{governato_10a, oh_11a, oh_11b}. Furthermore, outflows and winds in dwarf
galaxies may be behind the magnetization of the early Universe \citep[e.g.][]{pakmor_14a, siekowski_14a}.

Massive stars are the agents of stellar feedback and they manifest themselves
by carving bubbles -- cavities of tenuous, hot gas -- into the
ISM. They usually form in groups, so that their
bubbles start to overlap when expanding and subsequently merge, forming larger
structures in excess of 100~pc, so-called superbubbles. 
The wind of
massive stars, especially during their Wolf--Rayet (WR) phase,
powers the early expansion of the bubble. Subsequent SNe create
strong shocks in the bubble interior that are responsible for the thermal
X-ray and the non-thermal synchrotron emitting gas
\citep{krause_14a}. Stellar feedback in the form of SNe is more efficient for
clustered SNe than for randomly distributed ones as subsequent SNe explode in
the tenuous gas of the bubble and their shock waves are not suffering from
strong radiative cooling. Hence, the thermalization fraction
of clustered SNe is higher \citep{krause_13a}.

An intriguing example of SN feedback is presented by what has become known as the non-thermal superbubble \citep[NSB;][]{yang_93a} in the nearby
dwarf irregular galaxy IC~10, a member of the Local Group at a
distance of $0.7$~Mpc \citep[$\rm 1~arcsec = 3.4$~pc;][]{hunter_12a}. It has several young star clusters, containing massive
stars \citep{hunter_01a} and an unusually high number of WR stars \citep{massey_02a}. IC~10 is
a dwarf irregular galaxy that is currently undergoing a
starburst phase. Close to the centre of the NSB is one of the heaviest
stellar-mass black holes known at a remnant mass of $>$$23~\rm M_{\sun}$
\citep{silverman_08a}, which
forms together with the massive WR star [MAC92]~17A a
highly variable luminous X-ray binary, known as IC~10 \hbox{X-1} \citep[$\rm J2000.0$,
$\rm RA~00^h20^m29^s.09$, $\rm Dec.~59\degr 16\arcmin 51\farcs
95$;][]{bauer_04a,barnard_14a}.
 It has been speculated that a core collapse of the IC~10 X-1
progenitor in a `hypernova' could be responsible for the NSB,
rather than a series of SNe \citep{lozinskaya_07a}.

In this Letter, we present multiband radio continuum observations
with the NRAO\footnote{The National Radio Astronomy
  Observatory is a facility of the National Science Foundation operated under
  cooperative agreement by Associated Universities, Inc.} Karl G.\ Jansky Very Large Array (VLA) to study the non-thermal radio
continuum spectrum of the NSB. This project follows on, and extends some
preliminary results presented in \citet{heesen_11a}. The data cover the
frequency range between $1.4$ and $32$~GHz, at high spatial resolution.

\section{Observations}
We observed IC~10 with the VLA (project ID: AH1006). Observations
were taken in D-array in 2010 August and September at $L$ band ($1.4$--$1.6$~GHz),
$C$ band ($4.5$--$5.4$ and $6.9$--$7.8$~GHz), $X$ band ($7.9$--$8.8$~GHz) and
$Ka$ band (27--28 and 37--38~GHz) with $\approx$3~h on-source time each. In
addition, we observed in C-array at $L$, $C$ and $X$ band with
$\approx$3~h on-source time each in 2012 February to April (ID: 12A-288) and 2013
August (ID: 13A-328). A flux calibrator (3C~48) was observed either at the beginning or the end
of the observations, and scans of IC~10 were interleaved every 15~min
with a 2~min scan of a nearby complex gain calibrator (J0102+5824).  We incorporated
$L$-band B-array data observed with the historical VLA in 1986 September
(ID: AS0266) from \citet{yang_93a}.

We followed standard data reduction procedures, using the Common Astronomy
Software Applications package ({\small CASA}), developed by NRAO, and utilizing
the flux scale by \citet{perley_13a}.  We self-calibrated
the $L$-, $C$- and $X$-band data with two rounds of phase-only antenna-based gain corrections,
using images from the C-array data as a model. In $C$ and $X$ bands, we
self-calibrated in phase and amplitude, adding in the
D-array data (self-calibrated in phase), checking that the amplitudes did not
change by more than 1--2 per cent. For the imaging we used {\small CASA}'s implementation of
the Multi-Scale Multi-Frequency Synthesis (MS--MFS) algorithm \citep{rau_11a},
which simultaneously solves for spatial and spectral structure during
wide-band image reconstruction. A radio spectral index image was produced by
MS--MFS as well, which we used to refine the self-calibration model. A post-deconvolution wide-band primary beam correction
was applied to remove the effect of the frequency-dependent primary beam.  For
the spectral analysis, we imaged subsets (`spectral windows') of data with either 128 or 256~MHz
bandwidth, varying Briggs' `robust' parameter as function of
frequency to achieve a synthesized beam of a similar angular size. All data were
convolved with a Gaussian kernel in {\small AIPS}\footnote{{\small AIPS}, the Astronomical Image
  Processing Software, is free software available from the NRAO.} to an identical resolution
of $5.1$~arcsec and regridded.

Because the VLA cannot
record baselines smaller than $\approx$30~m (elevation dependent), there is a
limit to the largest angular scale that can be observed, resulting broadly in
flux densities that are too low compared with single-dish measurements; this
is known as the `missing zero-spacing flux'. Our VLA flux density at $1.5$
GHz and those at $2.6$ and $10.5$~GHz, measured with the 100-m
Effelsberg telescope \citep{chyzy_03a,chyzy_11a}, of 343, 277 and 156~mJy, can
be fitted with a
constant spectral index of $-0.41$. We can interpolate them to
estimate the missing zero-spacing flux in each spectral window. We found that
at frequencies of 4--6~GHz, 10--20 per cent of the flux
density was missed by the VLA, which increased to 30--40 per cent at
frequencies of 7--9~GHz. In order to correct for this,
we merged the VLA and Effelsberg data using {\small IMERG} in {\small AIPS}.  We used
the VLA $1.5$~GHz map as a template for
the large-scale emission at the lower end of our frequency range, which has the
benefit of an
improved angular resolution in comparison to the $2.6$~GHz Effelsberg map. We hence interpolated the
$1.5$~GHz VLA and $10.5$~GHz Effelsberg maps at an angular resolution of
78~arcsec, assuming a constant but spatially resolved spectral index, to have
a template of the large-scale emission in each spectral window.  We merged the data of
each spectral window with the appropriate template,  making sure that the integrated
flux densities of several regions agreed to within 5--10 per cent and the
discrepancy between the total integrated flux densities was less than 4
per cent. This was achieved by adjusting
the `uvrange' parameter in {\small IMERG}, which prescribes the angular scale at which the
single-dish image is scaled to interferometric image; we used values within the range of $0.8$--$1.8~\rm k\lambda$.
\begin{figure*}
  \includegraphics[width=1.0\hsize]{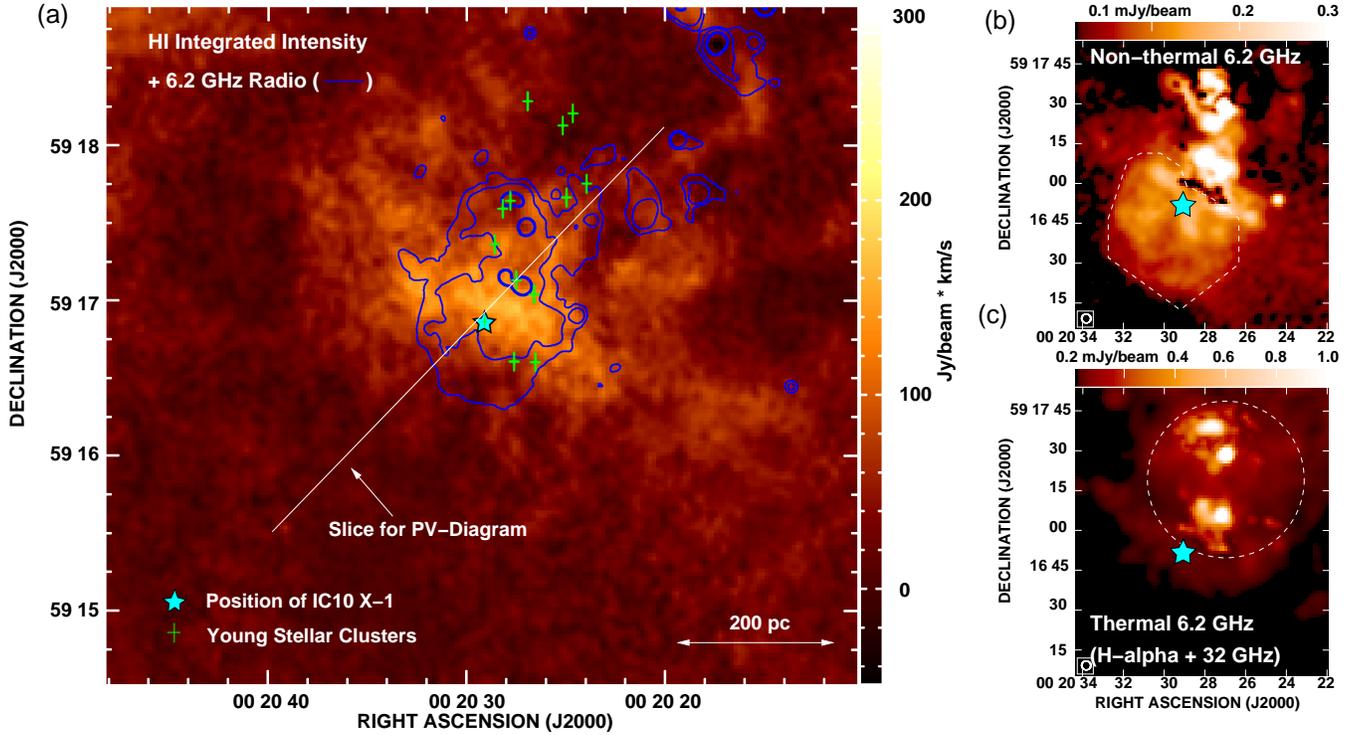}
\caption{(a) Integrated \HI emission line intensity as grey-scale at
  $5.5\times 5.9$~arcsec ($\rm PA=16\degr$) resolution of an approximately
  1~$\rm kpc^2$ region to the south-east of the centre of IC~10. Contours show the $6.2$~GHz
  radio continuum emission at 60, 120 and 800~$\rm\mu Jy\,beam^{-1}$, i.e.\ the superposition of thermal and non-thermal emission. The white line corresponds to
  the
  slice used to extract the $PV$-diagram at an angle of $-45\degr$, centred on the H\,{\small
  I} hole (see the text for details). Green plus signs show the position of
  stellar clusters \citep{hunter_01a}. (b) Non-thermal radio continuum at
  $6.2$~GHz of the superbubble, where the dashed line indicates the region, used
  for measuring the spectrum of the non-thermal superbubble. (c) Thermal radio continuum of the
same region as in (b), constructed from a combination of H$\alpha$
and 32~GHz emission. The dashed line indicates
 the 80~per cent attenuation level of the
primary beam at 32~GHz. In panels (a)--(c), the magenta star indicates the position of
IC~10 \hbox{X-1} and the angular resolution of the radio data is $3.4$~arcsec
(equivalent to 12~pc).}
\label{fig:6cm}
\end{figure*}

\section{Results}

In Fig.~\ref{fig:6cm}(a), we present a $6.2$~GHz contour map from combined VLA and Effelsberg observations
at $3.4$~arcsec angular resolution, overlaid on an integrated \HI
map from LITTLE THINGS \citep{hunter_12a}. The NSB is centred on $\rm
RA~00^h20^m28^s.85$, $\rm Dec.~59\degr 16\arcmin 48\arcsec $, which is 5~arcsec
south-west of IC~10 \hbox{X-1}, and has a
diameter of 54~arcsec or 184~pc. We created a map of the thermal (free--free)
emission from the Balmer H$\alpha$ emission map of
\citet{hunter_04a} following standard conversion
\citep[e.g.,][equation~3, $T=10^4$~K]{deeg_97a}, where we corrected for foreground absorption using $E(B-V)
= 0.75$~mag \citep{burstein_84a}. This map was combined with our $32$~GHz map
of the south-eastern starburst region,
which we use as an extinction free measurement of the thermal radio continuum
emission (Fig.~\ref{fig:6cm}~c). A
comparison between the two maps showed
agreement to within 10--20~per cent in areas outside of
the compact H\,{\small II} regions ($I_{\rm th}<1.0~\rm mJy\,beam^{-1}$), indicating that our estimate of the optical foreground
absorption is accurate.

\begin{figure}
  \includegraphics[width=0.95\hsize]{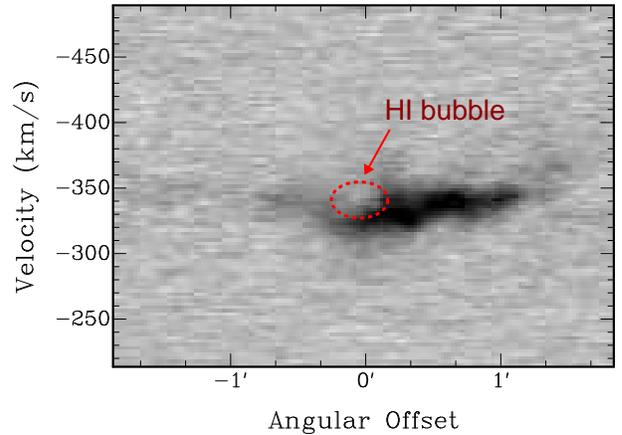}
\caption{Position--velocity ($PV$) diagram of the NSB and its
  surroundings, from the LITTLE THINGS
  \HI data cube. The position of the slice is shown in
  Fig.~\ref{fig:6cm}(a). South-east is to the left, north-west to the right.}
\label{fig:pv}
\end{figure}

The main
fraction of thermal radio continuum is located in the H\,{\small II} regions, north-west of
the NSB. Whereas the NSB is prominent in the non-thermal radio continuum,
there has thus far been little other evidence reported in the literature that the NSB constitutes
a cavity in the ISM. \citet{wilcots_98a} find an \HI hole at its
position, but do not report any signs of expansion. There is weak, diffuse
H$\alpha$ emission from ionized hydrogen and an
increased line width, corresponding to a thermal velocity dispersion of
$35~\rm km\,s^{-1}$, but nothing to suggest an expanding
shell \citep{thurow_05a}. Using the
natural weighted \HI data cube from LITTLE THINGS \citep[$\rm
FWHM=8.4\times 7.5$~arcsec, $\rm PA=37\degr$;][]{hunter_12a}, we have
created a position--velocity diagram of the NSB and its surroundings,
presented in Fig.~\ref{fig:pv}. We find a cavity of little
prominence, centred on $\rm RA~00^h20^m29^s.71$, $\rm Dec.~59\degr
16\arcmin 51\farcs 9$, which is $4.8$~arcsec east of IC~10 \hbox{X-1}. It is either a
single \HI hole with a
diameter of 100~pc and an
extent in velocity space of 30~$\rm km\,s^{-1}$, or consists of two smaller
\HI holes with diameters of 76~pc and extents in velocity space of
18~$\rm km\,s^{-1}$ each. For a single hole the expansion
velocity is 15~$\rm km\,s^{-1}$, leading to an estimate of the
bubble's dynamical age of $\tau_{\rm dyn}=3.8\pm 0.5$~Myr, with a similar age
for the double-hole scenario.

\begin{figure}
  \includegraphics[width=0.95\hsize]{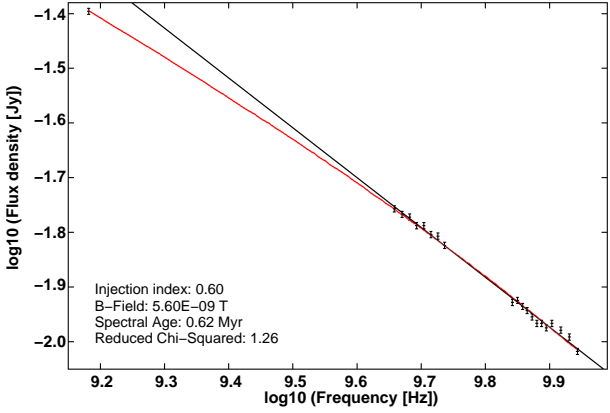}
\caption{Non-thermal spectrum of the NSB between $1.5$ and
$8.8$~GHz. The solid red line shows the Jaffe--Perola model fit to the data and the
solid black line is a linear fit to data points $>$$1.5$~GHz.}
\label{fig:spec}
\end{figure}

For further analysis, we fed the radio continuum data into the Broadband Radio Analysis ToolS
\citep[{\small BRATS};][]{harwood_13a}. The
spectrum of the NSB presented in Fig.~\ref{fig:spec} (flux densities are tabulated
in Table~\ref{tab:flux}) shows a conspicuous curvature, which can be well fitted by a
Jaffe--Perola \citep[JP;][]{jaffe_73a} model, shown as red solid line, with an
injection spectral index of $\alpha_{\rm inj}=0.6\pm 0.1$. The JP
model describes the evolution of radio continuum emission from a cosmic ray
electron (CRe) population within a constant magnetic field
strength following a single-injection. There exist variations to the JP
model, such as the KP \citep{kardashev_62a} and Tribble \citep{tribble_93a}
model. Our data cannot differentiate between them as any differences are only notable close to the break frequency. Assuming energy
equipartition and using the revised equipartition formula by \citet{beck_05a}, we find a
total magnetic field strength of $44\pm 8~\mu\rm G$ ($U_{\rm B}=7.7\pm 0.4 \times 10^{-11}~\rm erg\,cm^{-3}$). The total infrared luminosity from
\emph{Spitzer} MIPS 24--$160~\mu\rm m$ maps from \citet{dale_02a}
lead to a radiation energy density of $U_{\rm rad}=U_{\rm star}+U_{\rm IR} = 5\pm 1
\times 10^{-11}~\rm erg\,cm^{-3}$, where the contribution from stellar
light is taken as $U_{\rm star}=1.73\times U_{\rm IR}$ as measured in the solar
neighbourhood \citep{draine_11a}.

\begin{table}
\centering
\caption{Non-thermal flux densities of the NSB.\label{tab:flux}}
\begin{tabular}{lcccccc}
\hline\hline
$\nu$ (GHz) & $S_\nu$ (mJy) & $\nu$ (GHz) & $S_\nu$ (mJy) &$\nu$ (GHz) & $S_\nu$ (mJy)\\
\hline
$1.52$  & $40.2$  & $5.32$ & $15.6$ & $7.59$  & $10.8$\\
$4.55$  & $17.5$  & $5.45$ & $15.0$ & $7.72$  & $10.8$\\
$4.68$  & $17.1$  & $6.95$ & $11.8$ & $7.85$  & $10.6$\\
$4.81$  & $16.9$  & $7.08$ & $11.9$ & $8.01$  & $10.8$\\
$4.93$  & $16.3$  & $7.21$ & $11.6$ & $8.27$  & $10.5$\\
$5.06$  & $16.3$  & $7.33$ & $11.4$ & $8.53$  & $10.2$\\
$5.19$  & $15.7$  & $7.46$ & $11.1$ & $8.78$  & $ 9.6 $\\
\hline
\end{tabular}
\end{table}

\begin{figure}
  \includegraphics[width=0.95\hsize]{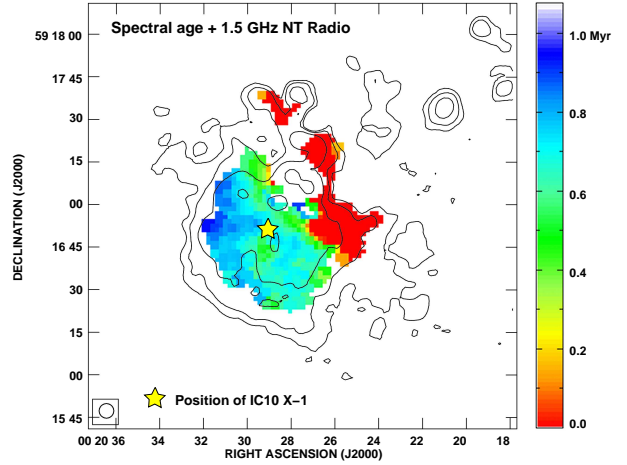}
\caption{Spectral age of the cosmic ray electrons
  in the NSB and its environment. The angular resolution is $5.1$~arcsec (equivalent to 17~pc) as indicated by the boxed circle in the bottom-left
corner. Contours show the non-thermal $1.5$~GHz emission at 3, 5, 10, 20 and
40 $\times$ 27~$\mu$Jy\,beam$^{-1}$ and the yellow star the position of
IC~10 \hbox{X-1}.}
\label{fig:age}
\end{figure}

The
spatially resolved distribution of the spectral age is shown in
Fig.~\ref{fig:age}, where we applied a S/N-cutoff of 5 in each pixel. There is an east--west gradient, where the age in the eastern part is
$\tau= 1.0$~Myr. The JP model fit to the spatially
resolved data is better ($\langle\chi^2_{\rm red}\rangle= 0.6$) than for the integrated
spectrum ($\chi^2_{\rm red}=1.3$), because a
superposition of spectral ages cannot be described by a single JP
model. This leads us to conclude that our best estimate of the spectral age is
$\tau_{\rm spec}=1.0\pm 0.3$~Myr. The error is a combination of the magnetic field error and
the formal fit error of the JP model.

Finally, in order to exclude a power-law
spectrum we conducted a few more
tests: first, we fitted a power-law fit to the integrated spectrum and found
$\chi_{\rm red}^2=4.5$, far inferior to the JP model fit. Secondly, a power-law fit to data points $>$$1.5$~GHz
predicts a flux density at $1.5$~GHz of 19 per cent above to the actual value,
about 15 times larger than its error of $1.3$ per cent (made up of 1~per cent
flux calibration error and the rms map noise) as shown in Fig.~\ref{fig:spec}. Thus, we can exclude a spectrum without
curvature.

\section{Discussion}
We first review the parameters derived for the IC~10 NSB:
the integrated current cosmic ray energy in the NSB is $1\times
  10^{51}$~erg, 
where we modelled the bubble as a sphere and used the
  assumption of energy equipartition ($U_{\rm CR}=U_{\rm B}$), injected
  approximately 1 ~Myr ago.
 Following the calculations of
\citet{bagetakos_11a}, we can derive the energy required to create the
\HI hole as  $0.2$--$1\times 10^{51}$~erg, with the upper value
appropriate if two holes were formed, where the ambient density of the
neutral, atomic
gas is $0.8$--$2.2~\rm cm^{-3}$ (including helium).  The contribution from turbulent gas
within the NSB traced in H$\alpha$ \citep{thurow_05a} is probably not significant
when taking into account that the filling factor for emission line gas is probably low.

We can compare our findings with 3D simulations by
\citet{krause_13a,krause_14a}. They predict that superbubbles 
reach diameters of the order of 100~pc even before the first SN. Each SN 
then first heats the bubble, accelerates the shell, and then dissipates the injected energy 
entirely at the leading radiative shock wave, and via radiative cooling in mixing regions 
at the location of the shell, on a time-scale of a few $10^5$~yr. The shell slows down
accordingly, resulting in a discrepancy between spectral and dynamical age. The rather low shell velocity 
of the IC~10 NSB
\citep[high-velocity superbubbles have a few times faster
shells, compare e.g.][]{Oey96} is indeed expected if the last embedded SN 
exploded about 1~Myr ago, as suggested by the non-thermal emission.
The CRe would have been accelerated as the accompanying shock wave 
traversed the bubble. 
Using the method of \citet{bagetakos_11a} on the aforementioned 3D~simulations 
at a similar time, 
we find $10^{51}$~erg as minimum energy to create the cavity, in agreement
with the upper limit from the observations. The energy found in cosmic rays is however 
surprisingly large. Assuming an acceleration efficiency of 10~per~cent
\citep[e.g.][]{Riegea13}, at least $10^{52}$~erg would have to have been released.

Could this have happened in a single explosion? Highly energetic SNe are thought to be related to long duration gamma-ray bursts \citep[e.g.][and references therein]{Mazea14}.
The associated Type~Ic SNe have energies of up to a few times $10^{52}$~erg,
adequate to account for our observations. We note that a higher energy than the standard $10^{51}$~erg would also better explain the high shell velocities in 
some high-velocity superbubbles \citep{Oey96,KD14}. It is noteworthy that the NSB is centred to within 16~pc on IC~10 \hbox{X-1},
suggesting an association. This system contains at least one massive star,
[MAC92] 14A, which has a mass larger than $17~\rm M_{\sun}$ and more likely
$35~\rm M_{\sun}$ \citep{silverman_08a}, also a possible progenitor for a Type Ic
SN. Alternatively, multiple SNe may have exploded in the past 1~Myr. We cannot
rule this out from the spectral ageing analysis, because a constant CRe
injection rate since approximately 1~Myr would still lead to a spectral
downturn, caused by the oldest CRe. However, the position of the
stellar clusters (Fig.~\ref{fig:6cm}~a) and the distribution of the
thermal radio continuum emission and hence that of massive stars
(Fig.~\ref{fig:6cm}~c), argues against this scenario, because there is no spatial
correlation. It is, however, conceivable that a less massive SN has
  exploded more recently, offset from IC~10 \hbox{X-1}, which could explain the
  east--west gradient in the spectral age distribution.

Another way to explain the presence of non-thermal particles would
 be
perhaps the energy release from IC~10 \hbox{X-1}. It is a debated possibility that
the X-ray emission of black hole binaries partially originates
from a jet in addition to that of the more conventional X-ray corona \citep{Grinea14}.
If the
current X-ray luminosity of IC~10 \hbox{X-1}, $10^{39}$~erg~s$^{-1}$ \citep{barnard_14a}, comes
exclusively from the jet, an outburst length
of 1~Myr would be sufficient to explain the cosmic ray energy, assuming a
10~per~cent acceleration efficiency. One
would then have to explain why this channel was so active in the past and by
now has ceased almost entirely with no compact radio source observed in the
vicinity of IC~10 \hbox{X-1}.

\section{Conclusions}

In this Letter, we have presented a multiband radio continuum study of the
NSB in the nearby starburst dwarf irregular galaxy IC~10. Conventional
wisdom tells us that dwarf galaxies are weak in non-thermal
(synchrotron) emission, being easily subjected to outflows and winds and
not likely able to retain cosmic rays. IC~10 is no exception,
it has a large thermal fraction of
50~per cent at 6~GHz and is underluminous in terms of its radio continuum
emission compared to its
star-formation rate \citep{heesen_11a,heesen_14b}. However, high spatial resolution
observations (10--20~pc) show
complex cosmic ray and magnetic field distributions. The NSB stands
out as the brightest non-thermal structure and its spectrum
shows a conspicuous downturn towards higher frequencies, something that
to date has rarely been observed in any nearby galaxy.

We fit a JP spectral model to the data, which describes the radio continuum
emission of an ageing population of CRe in a constant magnetic
field. Estimating the magnetic field from equipartition and the radiation energy density from
\emph{Spitzer} MIPS maps, we find a spectral age of $\tau_{\rm spec}=1.0\pm 0.3$~Myr
(uncertainty stems from the errors of the magnetic field and the spectral
fitting errors). The bubble's dynamical age is $\tau_{\rm
  dyn}=3.8\pm 0.3$, measured from the expansion speed of its corresponding `\HI hole'. Our
results suggest that the NSB was generated by the wind of the progenitor
of IC~10 \hbox{X-1}, a massive stellar mass black hole, during its main-sequence life
and WR phase. 
Considering alternative explanations, we find that most likely a 
single energetic explosion of the progenitor of IC~10 \hbox{X-1} released $\gtrsim10^{52}$~erg,
accelerating the non-thermal particles and the shell at the same time. The latter
than slowed down via interaction with the ambient medium to its current
velocity of 15~km~s$^{-1}$. We are
observing the NSB in the early stages of its evolution, of what may become
over the next few 10--50~Myr a superbubble of a few hundred parsec size visible as
a large \HI hole.

\section*{Acknowledegements}
VH acknowledges support from the Science and Technology Facilities
  Council (STFC) under grant ST/J001600/1. MK acknowledges support by the DFG
  cluster of excellence `Origin and Structure of the Universe' and by the ISSI
  project `Massive star clusters across the Hubble time'. JJH wishes to thank the University of Hertfordshire for generous financial
  support and STFC for a STEP award. We thank our referee, Biman Nath, for a
  constructive and thoughtful report.

\bibliographystyle{mn2e}

{\small  \bibliography{bub}
}

{\small This paper has been typeset from a \TeX/\LaTeX file prepared by the author.}

\end{document}